\documentclass{article}
\usepackage{ismir,amsmath,cite}
\usepackage{graphicx}
\usepackage{color}
\usepackage{makecell}
\usepackage{url}
\usepackage{hhline} 
\usepackage{xcolor,colortbl}
\usepackage{color, colortbl}
\definecolor{Gray}{gray}{0.9}
\definecolor{LightCyan}{rgb}{0.88,1,1}
\usepackage{multirow}

\title{Automatic tagging using\\deep convolutional neural networks}
\oneauthor
{Keunwoo Choi, Gy\"orgy Fazekas, Mark Sandler}
{Queen Mary University of London \\ {\tt \{keunwoo.choi, g.fazekas, mark.sandler\}@qmul.ac.uk}}

\begin{document}
\maketitle
\begin{abstract}
We present a content-based automatic music tagging algorithm using fully convolutional neural networks (FCNs). We evaluate different architectures consisting of 2D convolutional layers and subsampling layers only. In the experiments, we measure the AUC-ROC scores of the architectures with different complexities and input types using the \textit{MagnaTagATune} dataset, where a 4-layer architecture shows state-of-the-art performance with mel-spectrogram input. Furthermore, we evaluated the performances of the architectures with varying the number of layers on a larger dataset (\textit{Million Song Dataset}), and found that deeper models outperformed the 4-layer architecture. The experiments show that mel-spectrogram is an effective time-frequency representation for automatic tagging and that more complex models benefit from more training data. 
\end{abstract}
\section{Introduction}\label{sec:introduction}
Music tags are a set of \textit{descriptive keywords} that convey high-level information about a music clip, such as emotion (sad, anger, happy), genre (jazz, classical) and instrumentation (guitar, strings, vocal, instrumental). Since tags provide high-level information from the listeners' perspectives, they can be used for music discovery and recommendation. 
 
Automatic tagging is a classification task that aims to predict music tags using the audio signal. This requires extracting acoustic features that are good estimators of the type of tags we are interested in, followed by a single or multi-label classification or in some cases, regression stage.
From the perspective of feature extraction, there have been two main types of systems proposed in the literature. Conventionally, feature extraction relies on a signal processing front-end in order to compute relevant features from time or frequency domain audio representation. The features are then used as input to the machine learning stage. However, it is difficult to know what features are relevant to the task at hand. Although feature selection have been widely used to solve this problem \cite{yaslan2006audio}, clear recommendations which provide good association of features with tag categories are yet to emerge. A more recent approach unifies feature extraction with machine learning to allow relevant features to be learnt automatically. This approach is known as feature learning and requires deep neural networks (DNNs). 

 Aggregating hand-crafted features for music tagging was introduced in \cite{tzanetakis2002musical}. Several subsequent works rely on a \textit{Bag of frames} approach - where a collection of features are computed for each frame and then statistically aggregated. Typical features are designed to represent physical or perceived aspects of sound and include MFCCs, MFCC derivatives, and spectral features (e.g. spectral roll-off and centroids).  
 Since these are frame-level features, their statistics such as mean and variance are computed \cite{tzanetakis2002musical}, or they are clustered and vector quantised \cite{liangcontent} to obtain clip-level features. Finally, classifiers such as k-NN or Support Vector Machines are applied to predict tags.


As alternative to the above systems, DNNs have recently become widely used in audio analysis, following their success in computer vision, speech recognition \cite{sainath2013deep} and auto-tagging \cite{hamel2011temporal, dieleman2014end, nam2015deep, van2014transfer}. From an engineering perspective, DNNs sidestep the problem of creating or finding audio features relevant to a task. Their general structure includes multiple hidden layers with hidden units trained to represent some underlying structure in data. 


%
In computer vision, deep convolutional neural networks (CNNs) have been introduced because they can simulate the behaviour of the human vision system and learn hierarchical features, allowing object local invariance and robustness to translation and distortion in the model \cite{lecun1995convolutional}. CNNs have been introduced in audio-based problems for similar reasons, showing state-of-the-art performance in speech recognition \cite{sainath2013deep} and music segmentation \cite{ullrich2014boundary}. 

Several DNN-related algorithms have been proposed for automatic music tagging too. In \cite{dieleman2013multiscale} and \cite{van2014transfer}, spherical k-means and multi-layer perceptrons are used as feature extractor and classifier respectively. Multi-resolution spectrograms are used in \cite{dieleman2013multiscale} to leverage the information in the audio signal on different time scales. In \cite{van2014transfer}, pre-trained weights of multilayer perceptrons are transferred in order to predict tags for other datasets. A two-layer convolutional network is used in \cite{dieleman2014end} with mel-spectrograms as well as raw audio signals as input features. In \cite{nam2015deep}, bag-of-features are extracted and input to stacked Restricted Boltzmann machines (RBM).

In this paper, we propose an automatic tagging algorithm based on deep \textit{Fully Convolutional Networks} (FCN). FCNs are deep convolutional networks that only consists of convolutional layers (and subsampling) without any fully-connected layer. An FCN maximises the advantages of convolutional networks. It reduces the number of parameters by sharing weights and makes the learned features invariant to the location on the time-frequency plane of spectrograms, i.e., it provides advantages over hand-crafted and statistically aggregated features by allowing the networks to model the temporal and harmonic structure of audio signals. In the proposed architecture, three to seven convolutional layers are employed combined with subsampling layers, resulting in reducing the size of feature maps to $1$$\times$$1$ and making the whole procedure fully convolutional. 2D convolutional kernels are then adopted to take the local harmonic relationships into account.

We introduce CNNs in detail in Section \ref{sec:CNN} and define the problem in Section \ref{sec:problem_definition}. Our architectures are explained in Section \ref{sec:architecture}, and their evaluation is presented in Section \ref{sec:experiments}, followed by conclusion in Section \ref{sec:conclusion}.

\vspace{-5pt}

\section{CNNs for Music Signal Analysis }\label{sec:CNN}
\vspace{-1pt}
\subsection{Motivation for using CNNs for audio analysis}

In this section, we review the properties of CNNs with respect to music signals. The development of CNNs was motivated by biological vision systems where information of local regions are repeatedly captured by many sensory cells and used to capture higher-level information \cite{lecun1995convolutional}. CNNs are therefore designed to provide a way of learning robust features that respond to certain visual objects with local, translation, and distortion invariances. These advantages often work well with audio signals too, although the topology of audio signals (or their 2D representations) is not the same as that of a visual image.


CNNs have been applied to various audio analysis tasks, mostly assuming that auditory events can be detected or recognised by \textit{seeing} their time-frequency representations. 
Although the advantage of deep learning is to learn the features, one should carefully design the architecture of the networks, considering to what extent the properties (e.g. invariances) are desired.

There are several reasons which justify the use CNNs in automatic tagging. First, music tags are often considered among the topmost high-level features representing song-level information above intermediate level features such as chords, beats, tonality and temporal envelopes which change over time and frequency. This hierarchy fits well with CNNs as it is designed to learn hierarchical features over multilayer structures. Second, the properties of CNNs such as translation, distortion, and local invariances can be useful to learn musical features when the target musical events that are relevant to tags can appear at any time or frequency range. 

\vspace{-3pt}
\subsection{Design of CNNs architectures}
There have been many variants of applying CNNs to audio signals. They differ by the types of input representations, convolution axes, sizes and numbers of convolutional kernels or subsamplings and the number of hidden layers.

\subsubsection{TF-representation}
Mel-spectrograms have been one of the widespread features for tagging \cite{dieleman2013multiscale}, boundary detection \cite{ullrich2014boundary}, onset detection \cite{schluter2014improved} and latent feature learning \cite{van2013deep}. The use of the mel-scale is supported by domain knowledge about the human auditory system \cite{moore2012introduction} and has been empirically proven by performance gains in various tasks \cite{van2013deep, dieleman2014end, nam2015deep, schluter2014improved, ullrich2014boundary}.
The Constant-Q transform (CQT) has been used predominantly where the fundamental frequencies of notes should be precisely identified, e.g. chord recognition \cite{humphrey2012rethinking} and transcription \cite{sigtia2015end}. 

The direct use of Short-time Fourier Transform (STFT) coefficients is preferred when an inverse transformation is necessary \cite{choiauralisation, simpson2015deep}. It has been used in boundary detection \cite{grill2015music} for example, but it is less popular in comparison to its ubiquitous use in digital signal processing. Compared to CQT, the frequency resolution of STFT is inadequate in the low frequency range to identify the fundamental frequency. On the contrary, STFT provides finer resolutions than mel-spectrograms in frequency bands$>$2kHz given the same number of spectral bands which may be desirable for some tasks. So far, however, it has not been the most favoured choice.

Most recently, there have been studies focusing on learning an optimised transformation from raw audio given a task. These are called end-to-end models and applied both for music \cite{dieleman2014end} and speech \cite{sainath2015learning}. The performance is comparable to the mel-spectrogram in speech recognition\cite{sainath2015learning}. It is also noteworthy that the learned filter banks in both \cite{dieleman2014end} and \cite{sainath2015learning} show similarities to the mel-scale, supporting the use of the known nonlinearity of the human auditory system. 

\subsubsection{Convolution - kernel sizes and axes}

Each convolution layer of size $H$$\times$$W$$\times$$D$ learns $D$ features of $H$$\times$$W$, where $H$ and $W$ refer to the height and the width of the learned kernels respectively. 
The kernel size determines the maximum size of a component it can precisely capture. If the kernel size is too small, the layer would fail to learn a meaningful representation of shape (or distribution) of the data. For this reason, relatively large-sized kernels such as $17$$\times$$5$ are proposed in \cite{humphrey2012rethinking}. This is also justified by the task (chord recognition) where a small change in the distribution along the frequency axis should yield different results and therefore frequency invariance shouldn't be allowed.

The use of large kernels may have two drawbacks however. First, it is known that the number of parameters per representation capacity increases as the size of kernel increases. For example, $5$$\times$$5$ convolution can be replaced with two stacked $3$$\times$$3$ convolutions, resulting in a fewer number of parameters. Second, large kernels do not allow invariance within its range. 

The convolution axes are another important aspect of convolution layers. For tagging, 1D convolution along the time axis is used in \cite{dieleman2014end} to learn the temporal distribution, assuming that different spectral band have different distributions and therefore features should be learned per frequency band. 
In this case, the global harmonic relationship is considered at the end of the convolution layers and fully-connected layers follow to capture it.
In contrast, 2D convolution can learn both temporal and spectral structures and has already been used in music transcription \cite{sigtia2015end}, onset detection \cite{schluter2014improved}, boundary detection \cite{ullrich2014boundary} and chord recognition \cite{humphrey2012rethinking}.


\subsubsection{Pooling - sizes and axes}\label{sec:kernel_size}

Pooling reduces the size of feature map with an operation, usually a \textit{max} function. It has been adopted by the majority of works that are relying on CNN structures. Essentially, pooling employs subsampling to reduce the size of feature map while preserving the information of \textit{an activation} in the region, rather than information about the whole input signal. 

This non-linear behaviour of subsampling also provides distortion and translation invariances by discarding the original location of the selected values. As a result, pooling size determines the \textit{tolerance} of the location variance within each layer and presents a trade-off between two aspects that affect network performance. If the pooling size is too small, the network does not have enough distortion invariance, if it is too large, the location of features may be missed when they are needed. In general, the pooling axes match the convolution axes, although it is not necessarily the case. What is more important to consider is the axis in which we need invariance. For example, time-axis pooling can be helpful for chord recognition, but it would hurt time-resolution in boundary detection methods.



\section{Problem Definition}\label{sec:problem_definition}

Automatic tagging is a \textit{multi-label classification task}, i.e., a clip can be tagged with multiple tags. It is different from other audio classification problems such as genre classification, which are often formalised as a single-label classification problem.
Given the same number of labels, the output space of multi-label classification can exponentially increase compared to single-label classification. Accordingly, multi-label classification tasks require more data, a model with larger capacity and efficient optimisation methods to solve. If there are $K$ exclusive labels, the classifier only needs to be able to predict one among $K$ different vectors, which are \textit{one-hot vectors}. With multiple labels however, the number of cases increases up to $2^K$.

In crowd-sourced music tag datasets \cite{bertin2011million,law2009evaluation}, most of the tags are \textit{false}(0) for most of the clips, which makes accuracy or mean square error inappropriate as a measure. Therefore we use the Area Under an ROC (Receiver Operating Characteristic) Curve abbreviated as AUC. This measure has two advantages. It is robust to unbalanced datasets and it provides a simple statistical summary of the performance in a single value. It is worth noting that a random guess is expected to score an AUC of $0.5$ while a perfect classification $1.0$, i.e., the effective range of AUC spans between [$0.5$, $1.0$].

\section{proposed architecture}\label{sec:architecture}

\begin{figure}[b]
 \centerline{
 \includegraphics[width=\columnwidth]{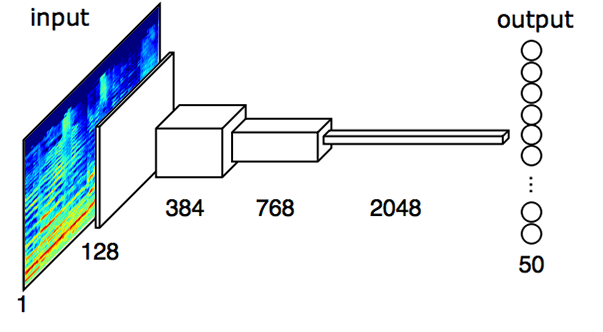}}
 \caption{A block diagram of the proposed 4-layer architecture, \textit{FCN-4}. The numbers indicate the number of feature maps (i.e. channels) in each layer. The subsampling layers decrease the size of feature maps to $1$$\times$$1$ while the convolutional layers increase the depth to $2048$.}
 \label{fig:example}
\end{figure}

Table \ref{table:architecture} and Figure \ref{fig:example} show one of the proposed architectures, a $4$-layer FCN (\textit{FCN-4}) which consists of $4$ convolutional layers and $4$ max-pooling layers. This network takes a log-amplitude mel-spectrogram sized $96$$\times$$1366$ as input and predicts a $50$ dimensional tag vector. The input shape follows the size of the mel-spectrograms as explained in Section~\ref{sec:exp_over}.

The architecture is extended to deeper ones with $5$, $6$ and $7$ layers (\textit{FCN-\{5, 6, 7\}}). The number of feature maps and subsampling sizes are summarised in Table~\ref{table:subsamples}. The number of feature maps of FCN-5 are adjusted based on FCN-4, making the hierarchy of the learned features deeper. FCN-6 and FCN-7 however have additional $1$$\times$$1$ convolutional layers(s) on the top of FCN-5. Here, the motivation of $1$$\times$$1$ is to take advantage of increased nonlinearity \cite{DBLP:journals/corr/LinCY13} in the final layer, assuming that the five layers of FCN-5 are sufficient to learn hierarchical features. An architecture with $3$ layers (FCN-3) is also tested as a baseline with a pooling strategy of [(3,5),(4,16),(8,17)] and [256, 768, 2048] feature maps. The number of feature maps are adjusted based on FCN-4 while the pooling sizes are set to increase in each layer so that low-level features can have sufficient resolutions.

\begin{table}[t]
\begin{center}
\begin{tabular}{ |c| }

\hline
FCN-4 \\
 \hline
 \hline

\rowcolor{Gray}
Mel-spectrogram \textit{(input: 96$\times$1366$\times$1)}\\
 \hline
 \hline
\makecell{Conv $3$$\times$$3$$\times$$128$}\\
 \hline
\makecell{MP ($2$, $4$) \textit{(output: 48$\times$341$\times$128)}}\\
 \hline
\makecell{Conv $3$$\times$$3$$\times$$384$}\\
 \hline
\makecell{MP ($4$, $5$) \textit{(output: 24$\times$85$\times$384)}}\\
 \hline
\makecell{Conv $3$$\times$$3$$\times$$768$}\\
 \hline
\makecell{MP ($3$, $8$) \textit{(output: 12$\times$21$\times$768)}}\\
 \hline
\makecell{Conv $3$$\times$$3$$\times$$2048$}\\
 \hline
\makecell{MP ($4$, $8$) \textit{(output: 1$\times$1$\times$2048)}}\\

 \hline
 \hline
\rowcolor{Gray}
Output 50$\times$1 (sigmoid)\\
 \hline
 \end{tabular}
\caption{The configuration of FCN-4}
\label{table:architecture}
\end{center}
\end{table}

Other configurations follow the current generic optimisation methods in CNNs. Rectified Linear Unit (ReLU) is used as an activation function in every convolutional layer except the output layer, which uses Sigmoid to squeeze the output within [0, 1]. Batch Normalisation is added after every convolution and before activation \cite{ioffe2015batch}. Dropout of 0.5 is added after every max-pooling layer \cite{srivastava2014dropout}. This accelerates the convergence while dropout prevents the network from overfitting.

Homogeneous 2D ($3$$\times$$3$) convolutional kernels are used in every convolutional layers except the final $1$$\times$$1$ convolution. 2D kernels are adopted in order to encourage the system to learn the \textit{local} spectral structures. The kernels at the first convolutional layer cover $64$ ms$\times$$72$ Hz. The coverage increases to $7$s$\times$$692$ Hz at the final $3$$\times$$3$ convolutional layer when the kernel is at the low-frequency. The time and frequency resolutions of feature maps become coarser as the max-pooling layer reduces their sizes, and finally a single value (in a $1$$\times$$1$ feature map) represents a feature of the whole signal.

\begin{table}[b!]
\begin{center}
\begin{tabular}{ |c|c|c| }
\hline
FCN-5 & FCN-6 & FCN-7 \\
 \hline
 \hline
\rowcolor{Gray}

 \multicolumn{3}{| c |}{
Mel-spectrogram \textit{(input: 96$\times$1366$\times$1)}}\\
 \hline
 \hline
 \multicolumn{3}{| c |}{ \makecell{Conv $3$$\times$$3$$\times$$128$}} \\
 \hline
  \multicolumn{3}{| c |}{ \makecell{MP ($2$, $4$) \textit{(output: 48$\times$341$\times$128)}} }\\
 \hline
  \multicolumn{3}{| c |}{ \makecell{Conv $3$$\times$$3$$\times$$256$}}\\
 \hline
  \multicolumn{3}{| c |}{ \makecell{MP ($2$, $4$) \textit{(output: 24$\times$85$\times$256)}} }\\
 \hline
  \multicolumn{3}{| c |}{ \makecell{Conv $3$$\times$$3$$\times$$512$}}\\
 \hline
  \multicolumn{3}{| c |}{ \makecell{MP ($2$, $4$) \textit{(output: 12$\times$21$\times$512)}}  }\\
 \hline
  \multicolumn{3}{| c |}{ \makecell{Conv $3$$\times$$3$$\times$$1024$} }\\
 \hline
  \multicolumn{3}{| c |}{ \makecell{MP ($3$, $5$) \textit{(output: 4$\times$4$\times$1024)}} }\\
 \hline
  \multicolumn{3}{| c |}{
 \makecell{Conv $3$$\times$$3$$\times$$2048$}}\\
 \hline
   \multicolumn{3}{| c |}{ \makecell{MP ($4$, $4$) \textit{(output: 1$\times$1$\times$2048)}} }\\
 \hline 
\multirow{2}{*}{\hphantom{conv 1} $\cdot$\hphantom{x1024}} & Conv $1$$\times$$1$$\times$$1024$ & Conv $1$$\times$$1$$\times$$1024$ \\
\cline{2-3}
 & $\cdot$ & Conv $1$$\times$$1$$\times$$1024$\\

 \hline
 \hline
\rowcolor{Gray}
 \multicolumn{3}{| c |}{
  Output 50$\times$1 (sigmoid)}\\
 \hline
 \end{tabular}
\caption{The configurations of 5, 6, and 7-layer architectures. The only differences are the number of additional 1$\times$1 convolution layers.}
\label{table:subsamples}
\end{center}
\end{table}

Several features in the proposed architecture are distinct from previous studies. Compared to \cite{van2014transfer} and \cite{nam2015deep}, the proposed system takes advantages of convolutional networks, which do not require any pre-training but fully trained in a supervised fashion. The architecture of \cite{dieleman2014end} may be the most similar to ours. It takes mel-spectrogram as input, uses two 1D convolutional layers and two (1D) max-pooling layers as feature extractor, and employs one fully-connected layer as classifier. The proposed architectures however consist of 2D convolution and pooling layers, to take the potential local harmonic structure into account. Results from many $3$s clips are averaged in \cite{dieleman2014end} to obtain the final prediction. The proposed model however takes the whole $29.1$s signal as input, incorporating a temporal nonlinear aggregation into the model.

The proposed architectures can be described as fully-convolutional networks (FCN) since they only consist of convolutional and subsampling layers. Conventional CNNs have been equipped with fully-connected layers at the end of convolutional layers, expecting each of them to perform as a feature extractor and classifier respectively. In general however, the fully connected layers account for the majority of parameters and therefore make the system prone to overfitting. This problem can be resolved by using FCNs with average-pooling at the final convolutional layer. For instance in \cite{DBLP:journals/corr/LinCY13}, the authors assume that the target visual objects may show large activations globally in the corresponding images. Our systems resemble the architecture in \cite{DBLP:journals/corr/LinCY13} except the pooling method, where we only use max-pooling because some of the features are found to be local, e.g. the voice may be active only for the last few seconds of a clip. 

\section{Experiments and Discussion}\label{sec:experiments}
\subsection{Overview}\label{sec:exp_over}
Two datasets were used to evaluate the proposed system, the MagnaTagATune dataset \cite{law2009evaluation} and the Million Song Dataset (MSD) \cite{bertin2011million}. The MagnaTagATune dataset has been relatively popular for content-based tagging, but similar performances from recent works \cite{nam2015deep, dieleman2014end, van2014transfer, dieleman2013multiscale} seem to suggest that performances are saturated, i.e. a glass-ceiling has been reached due to noise in the annotation. The MSD contains more songs than MagnaTagATune, it has various types of annotations up to 1M songs. There have not been many works to compare our approach with, partly because audio signals do not come with the dataset. Consequently, we use the MagnaTagATune dataset to compare the proposed system with previous methods and evaluate the variants of the system using the MSD. 

In Experiment I, we evaluate three architectures (FCN-\{3,4,5\}) with mel-spectrogram input as proposed in Section \ref{sec:architecture}. Furthermore, we evaluated STFT, MFCC, and mel-spectrogram representations as input of FCN-4. The architecture of STFT input is equivalent to that of mel-spectrograms with small differences in pooling sizes in the frequency axis due to the different number of spectral bands. For the architecture of MFCCs, we propose a frame-based $4$-layer feed-forward networks with time-axis pooling (instead of 2D convolutions and poolings) because relevant information is represented by each MFCC rather than its local relationships. In Experiment II, we evaluate five architectures (FCN-\{3,4,5,6,7\}) with mel-spectrogram input.

Computational cost is heavily affected by the size of the input layers which depends on basic signal parameters of the input data. A pilot experiment demonstrated similar performances with $12$ and $16$ kHz sampling rates and mel-bins of $96$ and $128$ respectively. As a result, the audio in both datasets was trimmed as $29.1$s clips (the shortest signal in the dataset) and was downsampled to $12$ kHz. The hop size was fixed at $256$ samples ($21$ ms) during time-frequency transformation, yielding $1$,$366$ frames in total. STFT was performed using $256$-point FFT while the number of mel-bands was set as $96$. For each frame, $30$ MFCCs and their first and second derivatives were computed and concatenated.

We used ADAM adaptive optimisation \cite{DBLP:journals/corr/KingmaB14} on Keras \cite{chollet2015} and Theano \cite{bastien2012theano} framework during the experiments. Binary cross-entropy function is used since it shows faster convergence and better performance than distance-based functions such as mean squared error and mean absolute error.

\vspace{-6pt}
\subsection{Experiment I: MagnaTagATune}\label{sec:exp1}

The MagnaTagATune dataset consists of $25$,$856$ clips of $29.1$-s, $16$ kHz-sampled mp3 files with $188$ tags. We only uses Top-$50$ tags, which includes genres (\textit{classical, rock}), instruments (\textit{piano, guitar, vocal, drums}), moods (\textit{soft, ambient}) and other descriptions (\textit{slow, Indian}). The dataset is not balanced, the most frequent tag is used $4$,$851$ times while the $50$-th most frequent one used $490$ times in the training set. The labels of the dataset consist of $7$,$644$ unique vectors in a $50$-dimensional binary vector space. 


\begin{table}[b]
\begin{center}
\begin{tabular}{ l|l }

\hline
Methods  & AUC \\ 
\hline
\hline
FCN-3, mel-spectrogram  & .852\\
  \hline
FCN-4, mel-spectrogram & \textbf{.894} \\ 
\hline

FCN-5, mel-spectrogram & .890\\
  \hline
FCN-4, STFT & .846\\
  \hline
FCN-4, MFCC & .862\\
  \hline

\end{tabular}
\caption{The results of the proposed architectures and input types on the MagnaTagATune Dataset}
\label{table:my_results}
\end{center}
\end{table}

\begin{figure}[b!]
 \centerline{
 \includegraphics[width=\columnwidth]{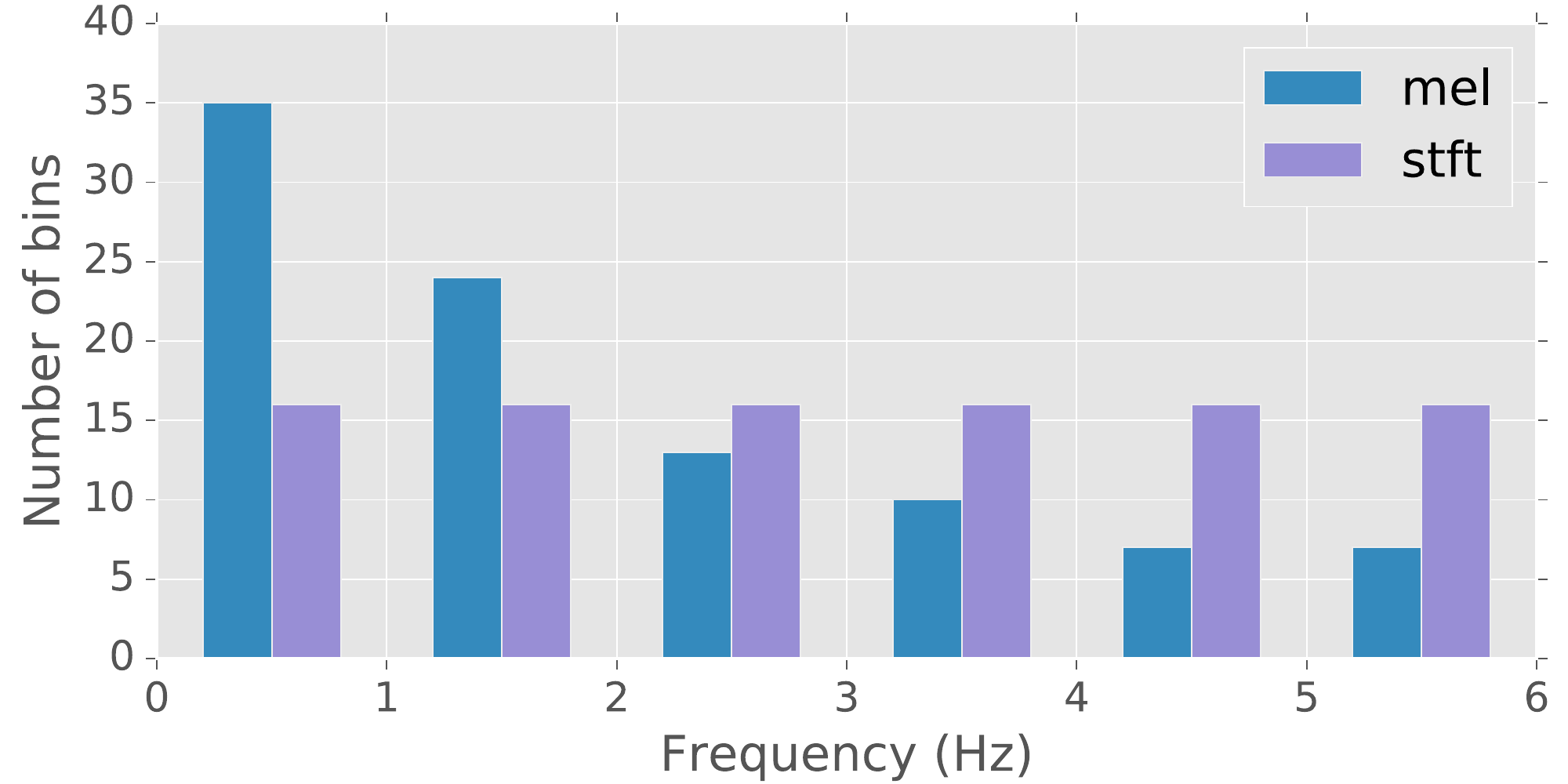}}
 \caption{The numbers of bins per 1kHz bandwidth in mel-spectrograms and STFTs .}
 \label{fig:num_bins}
\end{figure}

The results of the proposed architecture and its variants are summarised in Table \ref{table:my_results}. 
There is little performance difference between FCN-4 and FCN-5. It is a common phenomenon that an additional layer does not necessarily lead to an improved performance if, \textit{i}) the gradient may not flow well through the layers or \textit{ii}) the additional layer is simply not necessary in the task but only adds more parameters. This results in overfitting or hindering the optimisation. In our case, the most likely reason is the latter of the two. First, the scores are only slightly different, second, both FCN-4 and FCN-5 showed similar performances compared to previous research as shown in Table \ref{table:results}. Similar results were found in the comparison of FCN-5, FCN-6, and FCN-7 in Experiment II. These are discussed in Section \ref{sec:exp2}.

\begin{table}[t]
\begin{center}
\begin{tabular}{ l|l }
\hline
Methods  & AUC \\ \hline \hline
The proposed system, FCN-4 & .894 \\ \hline \hline
2015, Bag of features and RBM \cite{nam2015deep}  & .888\\   \hline
2014, 1D convolutions\cite{dieleman2014end} & .882\\  \hline
2014, Transferred learning \cite{van2014transfer} & .88\\  \hline
2012, Multi-scale approach \cite{dieleman2013multiscale} & .898\\  \hline
2011, Pooling MFCC \cite{hamel2011temporal} & .861\\  \hline
 \end{tabular}
\caption{The comparison of results from the proposed and the previous systems on the MagnaTagATune Dataset}
\label{table:results}
\end{center}
\end{table}


The degradations with other types of input signals--STFT and MFCC--are rather significant. This result is aligned with the preferences of mel-spectrograms over STFT on automatic tagging \cite{dieleman2013multiscale, nam2015deep, dieleman2014end, van2013deep}. However, this claim is limited to this or very similar tasks where the system is trained on labels such as genres, instruments, and moods. Figure \ref{fig:num_bins} shows how 96 frequency bins are allocated by mel-spectrograms and STFT in every 1kHz bandwidth. This figure, combined with the result in Table \ref{table:my_results} shows that high-resolution in the low-frequency range helps automatic tagging. It also supports the use of downsampling for automatic tagging. Focusing on low-frequency can be more efficient.


Table \ref{table:results} shows the performance of FCN-4 in comparison to the previous algorithms. The proposed algorithm performs competitively against the other approaches. However, many different algorithms only show small differences in the range of an AUC score of $0.88$ -- $0.89$, making their performances difficult to compare. This inspired the authors to execute a second experiment discussed in the next section.
In summary, the mel-spectrograms showed better performance than other types of inputs while FCN-4 and FCN-5 outperformed many previously reported architectures and configurations.

\subsection{Experiment II: Million Song Dataset}\label{sec:exp2}

We further evaluated the proposed structures using the Million Song Dataset (MSD) with \textit{last.fm} tags. We select the top 50 tags which include genres (\textit{rock, pop, jazz, funk}), eras (\textit{60s -- 00s}) and moods (\textit{sad, happy, chill}). 214,284 (201,680 for training and 12,605 for validation) and 25,940 clips are selected from the provided training/test sets by filtering out items without any top-50 tags. The number of tags ranges from 52,944 (\textit{rock}) to 1,257 (\textit{happy}) and there are 12,348 unique tag vectors. Note that the size of the MSD is more than 9 times larger than the MagnaTagATune dataset.

The results of the proposed architectures with different numbers of layers are summarised in Table \ref{table:msd_results}. Unlike the result from Experiment I, where FCN-4 and FCN-5 showed a slight difference of the performance (AUC difference of $0.008$), FCN-5,6,7 resulted in significant improvements compared to FCN-4, showing that deeper structures benefit more from sufficient data. 
However, FCN-6 outperformed FCN-5 only by AUC $0.003$ while FCN-7 even showed a slightly worse performance than FCN-6. This result agrees with a known insight in using deep neural networks. The structures of DNNs need to be designed for easier training when there are a larger number of layers \cite{he2015deep}. In theory, more complex structures can perform at least equal to simple ones by learning an identity mapping. Our results supports this. In the experiment, the performances of FCN-6 and FCN-7 were still making small improvements at the end of the training, implying it may perform equal to or even outperform FCN-5.
In practice, this approach is limited by computational resources and therefore \textit{very deep} structures may need to be designed to motivate efficient training, for instance, using deep residual networks \cite{he2015deep}.


\begin{table}[t]
\begin{center}
\begin{tabular}{ l|l }
\hline
Methods  & AUC \\ \hline \hline
FCN-3, mel-spectrogram  & .786 \\   \hline
FCN-4, --- &  .808\\  \hline
FCN-5, ---  & .848 \\  \hline
FCN-6, ---  & \textbf{.851}\\  \hline
FCN-7, ---  & .845\\  \hline
 \end{tabular}
\caption{The results from different architectures of the proposed system on the Million Song Dataset}
\label{table:msd_results}	
\end{center}
\end{table}

\begin{figure}[t!]
 \centerline{
 \includegraphics[width=\columnwidth]{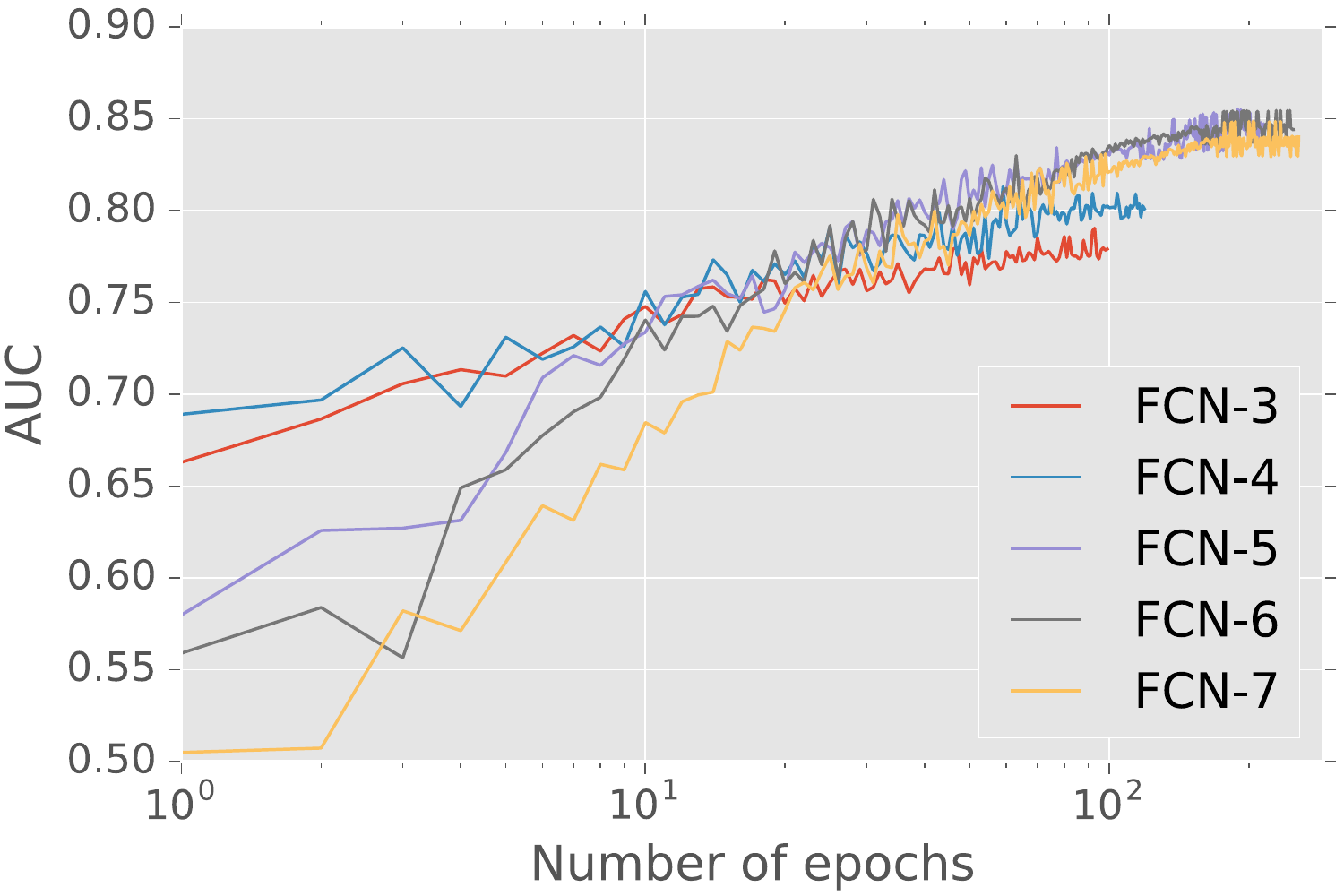}}
 \caption{The learning curves of the AUC scores measured on the validation set (on the Million Song Dataset)}
 \label{fig:curve}
\end{figure}

Figure \ref{fig:curve} illustrates the learning curves of the AUC scores on the validation set.  
At the beginning of the training, there is a tendency that simpler networks show better performance because there is a fewer number of parameters to learn. FCN-4 and FCN-5 show similar performance between around $20$--$40$ epochs. Based on this, it can be assumed that learning on the MagnaTagATune dataset stayed within this region and failed to make more progress due to the scarcity of training data.
To summarise, FCN-5, FCN-6, and FCN-7 significantly outperformed FCN-3 and FCN-4. The results imply that more complex models benefit from more training data. The similar results obtained using FCN-5, FCN-6 and FCN-7 indicate the need for more advanced design methodologies and training of deep neural networks.

\vspace{-5pt}
\section{Conclusion}\label{sec:conclusion}

We presented an automatic tagging algorithm based on deep fully convolutional neural networks (FCN). It was shown that deep FCN with 2D convolutions can be effectively used for automatic music tagging and classification tasks. In Experiment I (Section \ref{sec:exp1}), the proposed architectures with different input representations and numbers of layers were compared using the MagnaTagATune dataset against the results reported in previous works showing competitive performance. With respect to audio input representations, using mel-spectrograms resulted in better performance compared to STFTs and MFCCs. In Experiments II (Section \ref{sec:exp2}), different number of layers were evaluated using the Million Song Dataset which contains nine times as many music clips. The optimal number of layers were found to be different in this experiment indicating deeper networks benefit most from the availability of large training data. In the future, automatic tagging algorithms with variable input lengths will be investigated.

\vspace{-5pt}
\section{Acknowledgements}
This work was part funded by the FAST IMPACt EPSRC Grant EP/L019981/1 and the European Commission H2020 research and innovation grant AudioCommons (688382). Sandler acknowledges the support of the Royal Society as a recipient of a Wolfson Research Merit Award.

\vspace{-5pt}
\bibliography{tag_prediction_full_names}
\end{document}